\documentclass[runningheads]{llncs}
\usepackage[T1]{fontenc}

\usepackage[hyphens]{url}
\usepackage[hidelinks]{hyperref}
\usepackage{orcidlink}
\usepackage{comment}
\usepackage{xcolor}
\usepackage{cite}
\usepackage{graphicx}
\usepackage{textcomp}
\usepackage{amsmath}
\usepackage{amssymb}
\usepackage{mathrsfs}
\usepackage{epsfig}
\usepackage{epsf}
\usepackage{theorem}
\usepackage[active]{srcltx}
\usepackage{enumerate}
\usepackage{latexsym,calc,dsfont,array,pgf,setspace}
\usepackage{acro}
\usepackage{tikz}
\usetikzlibrary{calc}
\usepackage{pifont}
\usepackage{subfigure}
\usepackage{epstopdf}
\usepackage{amsfonts}
\usepackage{float}
\usepackage{xcolor}
\usepackage{afterpage}
\usepackage[inkscapeformat=png]{svg}
\usepackage{graphicx}
\usepackage{algorithm,algorithmic}
\usepackage[acronym,shortcuts]{glossaries}
\usepackage{soul}

\newacronym{owc}{OWC}{Optical Wireless Communication}
\newacronym{JCAS}{JCAS}{joint communication and sensing}
\newacronym{ISAC}{ISAC}{integrated sensing and communication}
\newacronym{mmWave}{mmWave}{millimeter-wave}
\newacronym{THz}{THz}{terahertz}
\newacronym{AI}{AI}{artificial intelligence}
\newacronym{V2V}{V2V}{vehicle to vehicle}
\newacronym{5G}{5G}{fifth-generation}
\newacronym{6G}{6G}{sixth-generation}
\newacronym{JRC}{JRC}{joint radar and communication}
\newacronym{JCR}{JCR}{joint communication and radar}
\newacronym{MIMO}{MIMO}{multiple-input and multiple-output}
\newacronym{AF}{AF}{ambiguity function}
\newacronym{SINR}{SINR}{signal to interference plus noise ratio}
\newacronym{LPI}{LPI}{low probability of intercept}
\newacronym{TDoA}{TDoA}{time difference of arrival}
\newacronym{AoA}{AoA}{angle of arrival}
\newacronym{AoD}{AoD}{angle of departure}
\newacronym{RSSI}{RSSI}{received signal strength indicator}
\newacronym{RDFC}{RDFC}{randomized distributed function computation}
\newacronym{CSI}{CSI}{channel state information}
\newacronym{D2D}{D2D}{device-to-device}
\newacronym{UE}{UE}{user equipment}
\newacronym{gdpr}{GDPR}{general data protection regulation}
\newacronym{QoS}{QoS}{quality-of-service}
\newacronym{MBB}{MBB}{mobile broadband}
\newacronym{URLLC}{URLLC}{ultra reliable low latency communications}
\newacronym{MMTC}{mMTC}{massive machine type communications}
\newacronym{IoS}{IoS}{internet of senses}
\newacronym{IoT}{IoT}{internet of things}
\newacronym{ITU}{ITU}{International Telecommunication Union}
\newacronym{ML}{ML}{Machine Learning}
\newacronym{LAAs}{LAAs}{large antenna arrays}
\newacronym{RIS}{RIS}{reconfigurable intelligent surfaces}
\newacronym{V2X}{V2X}{vehicle-to-everything}
\newacronym{AR}{AR}{augmented reality}
\newacronym{VR}{VR}{virtual reality}
\newacronym{PII}{PII}{personally identifiable information}
\newacronym{AN}{AN}{artificial noise}
\newacronym{CFR}{CFR}{Channel Frequency Response}
\newacronym{DoA-OPR}{DoA-OPR}{direction-of-arrival obfuscation power ratio}

\makeatletter
\def\namedlabel#1#2{\begingroup
    #2%
    \def\@currentlabel{#2}%
    \phantomsection\label{#1}\endgroup
}
\makeatother
    
\usepackage{graphicx}

\begin{document}
\title{ISAC Privacy: Challenges and Solutions for 6G}

\author{Onur Günlü\inst{1,2}\orcidlink{0000-0002-0313-7788} \and
Stefano Tomasin\inst{3}\and
João P.~Vilela \inst{4}\orcidlink{0000-0001-5805-1351}\and
Francesco Chiti \inst{5}\and
Prajnamaya Dass\inst{6}\and 
Angeliki Alexiou \inst{7}\and
Utz Roedig\inst{8}}

\authorrunning{O. Günlü et al.}

\institute{Lehrstuhl für Nachrichtentechnik, TU Dortmund, Germany \and
Information Theory and Security Laboratory (ITSL), Linköping University, Sweden \and 
Dept. of Information Eng. and Dept. of Mathematics, University of Padova, Italy \and 
INESCTEC, CISUC, and Dep.~of Computer Science, Faculty of Sciences, University of Porto, Portugal\and
Department of Information Engineering, University of Florence, Italy\and 
Barkhausen Institut, Dresden, Germany\and
Department of Digital Systems, University of Piraeus, Greece\and
School of Computer Science and Information Technology, University College Cork, Ireland\\
\email{onur.guenlue@tu-dortmund.de, stefano.tomasin@unipd.it, jvilela@fc.up.pt, francesco.chiti@unifi.it, prajnamaya.dass@barkhauseninstitut.org, alexiou@unipi.gr,  u.roedig@ucc.ie}}
\maketitle              
\begin{abstract}
Integrated sensing and communication (ISAC) is a promising feature of future communication networks. While spatial sensing can improve network performance and enable external services, it also creates privacy challenges that go beyond the confidentiality of communication content. Future networks using millimeter-wave (mmWave) and sub-terahertz (THz) frequencies may collect or infer detailed information about people, devices, bystanders, passive objects, and environments in a sixth-generation (6G) deployment area. Such sensing can reveal location and environment data, support behavioral profiling such as movement or activity recognition, and, in advanced cases, expose physiological information such as breathing frequency or heart-rate-related data. Thus, the capabilities of spatial sensing must be controlled to satisfy privacy requirements. In this work, we organize privacy-sensitive ISAC data into three sensing levels: location and environment data, behavioral data, and physiological data, and use this classification as the organizing principle throughout the paper. Based on this classification, we discuss internal and external ISAC applications, identify privacy challenges related to consent, transparency, data ownership, profiling, bystander exposure, and sensitive sensing data, review representative solution directions, and outline future research directions for privacy-preserving ISAC. 

\keywords{Integrated Sensing and Communication \and 6G Networks \and Privacy Preservation \and Privacy-aware Sensing Data \and User Profiling.}
\end{abstract}

\section{Introduction}\label{sec:introduction}
Integration of sensing and communication operations will be a key component of future communication networks~\cite{de2021convergent,GPFuKaVaDaShShByWy24}. Usually, spatial sensing is assumed when referring to the term \ac{ISAC}, also called \ac{JCAS}. Spatial sensing can be used within the network to improve performance, but it may also be used to provide external services. For instance, spatial sensing within the network can enable identification of temporary obstructions within a communication link, and the network could then take actions to react to the situation and change transmission arrangements. Similarly, external applications can make use of spatial sensing, such as traffic management systems that map the positions of vehicles and pedestrians~\cite{3GPP_usecase,LiGuHeMYuZhGu24}.

Spatial sensing via a network infrastructure is of particular interest due to the possibility of reusing parts of an already deployed communication infrastructure. Moreover, future networks are expected to use \ac{mmWave} and sub-\ac{THz} frequencies, large antenna arrays, and directional transmission, which can support more accurate sensing and localization~\cite{GaoTC2023,ZhHeYuLiHeTa24,GPFuKaVaDaShShByWy24}. Using such high frequencies requires beamforming to enable communication over a larger distance, enabling networks to scan an area using a narrow beam. Thus, a future communication network may provide many of the elements required to build a radar-like system with dense infrastructure and wide-area coverage. These same capabilities, however, also make it necessary to control when, how, and for which purpose sensing is performed. 

While implementing \ac{ISAC} systems enables network optimization and new applications, it also introduces new security and privacy challenges~\cite{wei2022toward,OnurSecureISAC,martins2025delving,qu2024privacy,OnurSecureCommunicationComputationTutorial}. Security challenges include deliberate manipulation of the sensing components of the network and attacks that degrade communication or sensing reliability. Besides security issues, the introduction of \ac{ISAC} creates significant privacy challenges. Indeed, \ac{ISAC} involves the collection and processing of sensing and situational data, including personally identifiable information; it poses major privacy risks~\cite{dass2024addressing,3GPP_usecase,qu2024privacy}. In this work, we focus on the privacy implications of sensing-derived personal and environmental information, and discuss ISAC privacy challenges, representative solution directions, and future research trends. Here, we define a privacy breach as the loss of control over information about a person, their device, or their surrounding environment that is revealed through sensing. Thus, disclosure of the {\em content} of a communication message is not considered a privacy issue in this paper, but rather a confidentiality and security issue. Moreover, a person affected by \ac{ISAC} sensing need not be an active network user, as passive objects and bystanders in the sensing area may also be included in sensing results~\cite{3GPP_usecase,dass2024addressing}. The discussion in this work covers sensing performed by base stations, cooperative network nodes, \acp{UE}, or combinations thereof.

Communication networks already gather large amounts of user information, such as user identifiers, traffic patterns, access times, and locations. However, when this information is combined with fine-grained spatial sensing information, a new level of privacy loss can occur. In this work, we organize these risks and the subsequent discussion according to three privacy-sensitive sensing-data levels: (i) location and environment data, (ii) behavioral data, and (iii) physiological data. Given the resolution expected from \ac{mmWave} and sub-\ac{THz} sensing, it can be possible to infer location and movement~\cite{10355064,WLAN_sensing}, behavior patterns such as gait, posture, gestures, or physical activity~\cite{CSITime,GaitRecon,Meneghello2023,Li2022,Ma2018}, and physiological parameters such as breathing frequency or heart-rate-related information~\cite{radar_biometric,ZhaoAccess2024,Ratnam2024}. Recent advances in \ac{AI} can further support the analysis of large sensing datasets and the creation of detailed profiles of people and environments~\cite{WuJiWaYaZhYiNa24,9146540}. Furthermore, it will be difficult for people to avoid being sensed (e.g., by not carrying a phone), because mere presence in an area with \ac{6G} coverage can subject them to sensing activity.   

As future \ac{6G} networks will be ubiquitous, it is necessary to design them with respect to user privacy. This is essential to ensure that users will trust and accept the deployment of networks that have the potential to infringe on their rights to privacy. To this end, the capabilities of spatial sensing may have to be balanced with privacy requirements such as sensitive data minimization, consent, transparency, purpose limitation, access control, and secure handling of personal data~\cite{gdpr,dass2024addressing,cigno2022integrating,report-CNIT}.

\begin{figure}[t]
    \centering
    \includegraphics[width=\linewidth]{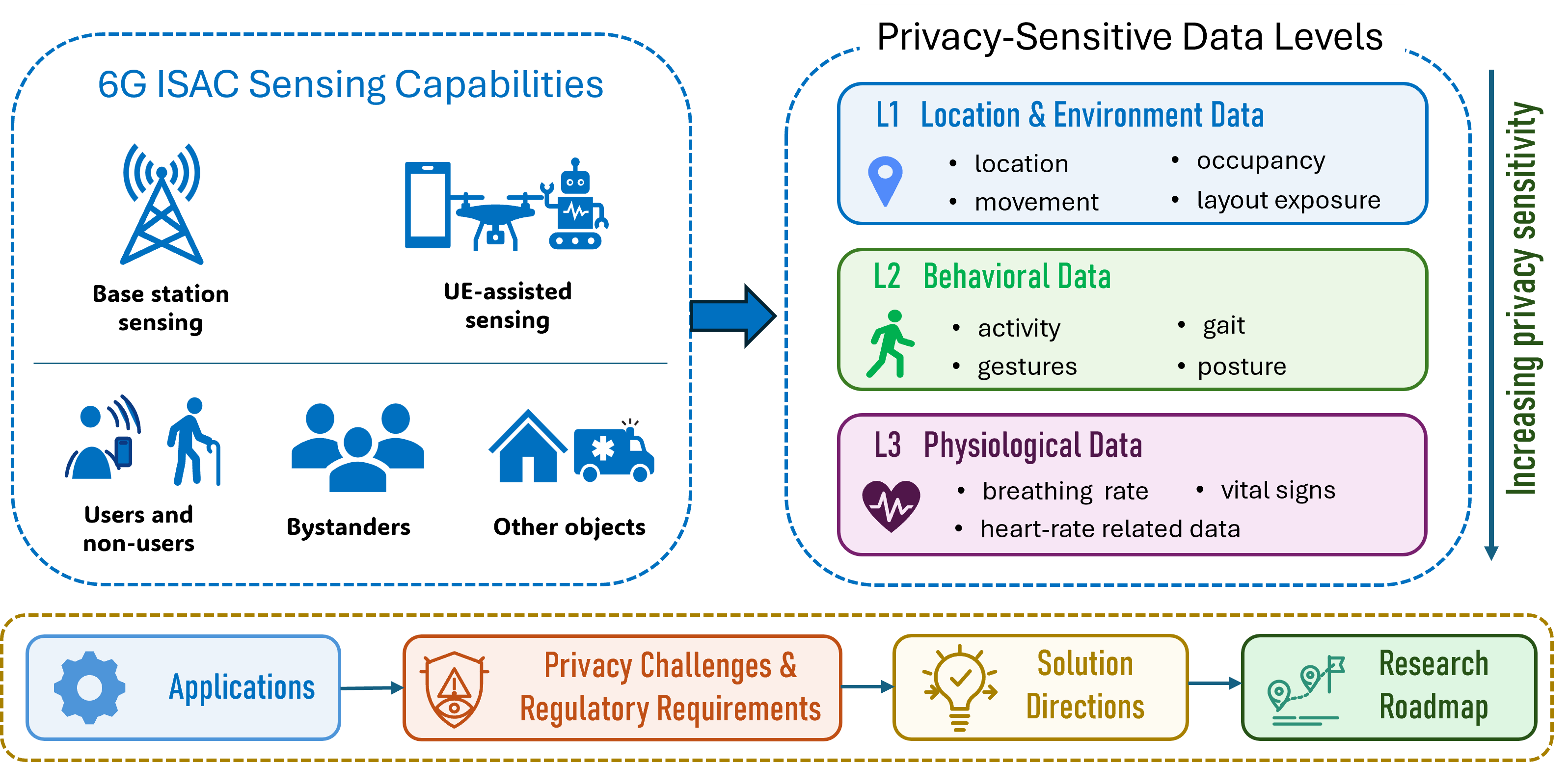}
    \vspace{-0.8cm}
    \caption{Conceptual overview of this work's perspective on \ac{6G} \ac{ISAC} privacy across base station-based and \ac{UE}-assisted sensing. The figure highlights that this paper focuses on sensing-derived personal data rather than message content, organizes privacy-sensitive \ac{ISAC} data into three levels (L1: location and environment data, L2: behavioral data, and L3: physiological data), and relates these levels to applications, privacy challenges, solution directions, and the research roadmap.}
    \label{fig:tutorial_overview}
\end{figure}

Recent tutorial and survey papers have covered \ac{ISAC} signal processing, applications, network architectures, secure \ac{ISAC}, and security/privacy challenges~\cite{LiuTC2020,ZhangJSTSP2021,JCAS_survey,GPFuKaVaDaShShByWy24,wei2022toward,10608156,martins2025delving,11358925,OnurSecureISACTutorial}. This work is complementary to those works, as it treats privacy as control over sensing-derived personal data and uses the three sensing-data levels introduced in Section~\ref{sec:level_sensing_granularity} as the organizing principle. Fig.~\ref{fig:tutorial_overview} provides a visual overview of this organizing principle.

In this work, we provide a brief overview of the envisioned \ac{ISAC} technology for \ac{6G} networks. Next, we outline applications for \ac{ISAC}, as it is essential to consider specific applications when considering privacy challenges. We then provide a structured discussion of privacy challenges, including consent, transparency, data ownership, location and environment data, behavioral data, and physiological data. This is followed by an overview of currently available solution directions. We also describe privacy-related research areas that must be addressed in the future to ensure that users will trust and accept \ac{ISAC}.

\section{Sensing in 6G Networks}
\label{sensing}
This section identifies the sensing capabilities needed to define the privacy-sensitive sensing-data levels used throughout this work.

\subsection{6G Networks}

The evolution of usage scenarios from \ac{5G} to \ac{6G}, as recommended by the \ac{ITU} in \cite{ITU_IMT2030_Framework_2024}, includes a group of extreme evolutions of the well-known \ac{5G} usage scenarios (\ac{MBB}, \ac{URLLC}, \ac{MMTC}) towards immersive and hyper reliable connectivity and a group of three revolutionary usage scenarios with a clear emphasis on ‘beyond communications’ objectives, such as ubiquitous coverage, artificial intelligence, and cyber-physical interaction with the \ac{IoS}.

The newly introduced usage scenarios, in particular ubiquitous connectivity, \ac{ISAC}, and integrated \ac{AI} and communication, introduce a major paradigm shift~\cite{ITU_IMT2030_Framework_2024,GPFuKaVaDaShShByWy24}. Connectivity in next-generation networks will not be only about transmission of bits: monitoring the environment, detecting and tracking mobility, localization, shape identification, and semantic understanding are among the targeted functionalities. Moreover, the adoption of \ac{AI}/\ac{ML} can support the analysis of complex and stochastic radio environments. Note that with \ac{ISAC} we refer to the reconstruction of the surrounding electromagnetic environment, in particular of the passive targets that are not active devices but objects (and people) hit by electromagnetic signals emitted by network devices (base stations or \acp{UE}). In this context, situational awareness, environmental sensing, tunability, and learning become central capabilities. More importantly, to assess such beyond-communication capabilities, \ac{6G} technical performance requirements need to go beyond traditional metrics such as data rate, spectral efficiency, latency, and reliability, and include sensing- and \ac{AI}-related metrics, positioning accuracy, energy efficiency, and resilience~\cite{ITU_IMT2030_Framework_2024,de2021convergent}. 

Defining quantitative measures for the new \ac{6G} capabilities is a challenging task, mainly because target values are hard to determine and are highly dependent on the specific use case and application. Even more important is the definition of the overarching objectives and strategic priorities in the design and deployment of \ac{6G} networks. As emphasized by the ITU, sustainability, "connecting the unconnected", and security and privacy are defining key values that \ac{6G} is expected to serve.

\subsection{Sensing Approaches}

In the \ac{6G} era, the conventional view of the network as a manager of physical and virtual resources is expected to evolve toward a system that supports programmability, power efficiency, resilience, intelligence, and sensing-as-a-service within an integrated communication-and-sensing architecture. Realizing such co-designed communication and sensing requires: (i) propagation and channel modeling principles, waveforms, signaling, protocols, and algorithms for \ac{ISAC}; (ii) sensing beyond conventional sensor/actuator and radar concepts, where communication systems also support beyond-communication functionalities; and (iii) integration of sensing technologies, \ac{ISAC}, and \ac{AI}/\ac{ML} capabilities into a wireless network optimization framework \cite{LiCuMaXuHaElBu22,LiuTC2020,ZhangJSTSP2021,JCAS_survey}.

Radio frequency sensing and wireless communications have traditionally been performed in separate systems, in distinct platforms and in non-cooperative settings~\cite{LiuTC2020,ZhangJSTSP2021}. They have different objectives and performance criteria, for example, data rate or quality of service in communications, and detection probability or the Cramer-Rao bound of target parameter estimation for sensing. Radio frequency sensing can be passive, as in localization via passive radars that may use communication signals as signals of opportunity, or active, where sensors probe their operational environment by emitting self-generated energy and observe the signals reflected off the targets as in active radars or radio frequency imaging. The sensing tasks involve target detection, tracking, target recognition, imaging, and channel and direction of arrival estimation, which are then directly used to aid communication tasks, such as data symbol estimation, or beamforming to enhance signal-to-noise ratio and mitigate interference. For the purpose of this paper, these sensing tasks are important because their outputs can become location, environment, behavioral, or physiological information rather than only communication-side state information.

In \ac{6G}, sensing will be used to obtain situational awareness in order to enhance communications performance~\cite{de2021convergent,GPFuKaVaDaShShByWy24,LiCuMaXuHaElBu22}. For example, by utilizing radio environment sensing, transceiver operational parameters can be optimized or estimated, and consequently, communication performance can be improved. For instance, knowledge of the locations of mobile users can simplify beam alignment and allow for managing interferences in the spatial domain. A key component of designing sensing-aided communications is to build and maintain awareness in a dynamic radio environment. Optimizing waveforms to be suitable for both sensing and communication tasks and incorporating adaptability and prediction through \ac{AI}/\ac{ML} have an important role to play. In addition to the incorporation of sensing and radar capabilities into communication systems and the exploitation of sensing information to enhance communication performance, \ac{6G} is anticipated to offer a revolutionary path towards perceptive intelligence by introducing the notion of network-(nodes)-as-a-sensor. The use of a single base station for sensing the environment and the cooperation of multiple network nodes (access points or remote radio heads) opens new ways to improve the sensing performance~\cite{GPFuKaVaDaShShByWy24,JCAS_survey}. Moreover, the mobile users may serve either as sensors or sources of radio signals to be exploited by the network to perform sensing. For example, cell-free network architectures, by exploiting signaling from multiple base stations and mobile stations, may enhance the sensing ability and provide sensing services, such as positioning of active users and mapping of the propagation environment, including blockage profiling. From a privacy perspective, this is important because sensing is no longer confined to dedicated radar devices or base stations, as cooperative network nodes and, in some settings, \acp{UE} may also contribute to the sensing process.

Motivated by the consideration of high frequency bands (\ac{mmWave} and sub-\ac{THz}), the adoption of \ac{LAAs} and large aperture \ac{RIS} promise attractive solutions for extreme spatial reuse, pencil beamforming, and transceiver tunability~\cite{GaoTC2023,ZhHeYuLiHeTa24,10639526}. Moreover, as the aperture of \ac{LAAs} increases, the Fraunhofer distance increases as well and they operate in the near field (spherical waves) for several scenarios of interest (at substantial distance from the transmitter). Near field communications with \ac{LAAs} offer the possibility to perform advanced wavefront engineering, involving phase shifts beyond the simple linear phase gradient that is required for conventional beam steering. For example, bending beams \cite{10791450} are able to propagate along a curved trajectory, while beams with self-healing properties are able to reconstruct after being interrupted by a small blocker. Focused beams, on the other hand, are ideal for controlling power density in small focal areas, which is a key feature for localization \cite{10838506}, mobility tracking and sensing applications \cite{10639526}.

\subsection{Privacy Sensitive Sensing Data}
\label{sec:level_sensing_granularity}

From the aforementioned overview on sensing technology for \ac{6G}, it is clear that privacy-sensitive information can be captured on different levels of granularity. We differentiate three levels of sensing granularity and, thus, resulting sensing data, linked to the capability of the present \ac{6G} deployment. The levels describe the type of information revealed, irrespective of whether the sensing data are generated by infrastructure nodes, \acp{UE} or both.
\begin{enumerate}
    \item[\namedlabel{itm:L1}{L1})] \emph{Location and Environment Data:} Tracking objects and people is the simplest (least technologically challenging) form of \ac{ISAC}. Such a system can provide information on user location and movement \cite{cunha2025compromising,11358925}, revealing other private information, such as the visited places and meetings among different users. It can also sense the surroundings, which may disclose other privacy-sensitive information, such as the floor plan of a residence.  
    \item[\namedlabel{itm:L2}{L2})]  \emph{Behavioral Data:} More technologically advanced sensing systems can also record behavioral information on users. For example, they can determine the user posture, distinguishing between sitting, standing, or lying down (sleeping) users \cite{11154859}. The movement of limbs may also be identified \cite{11153070}, and gait can be characterized, not only for users' identification but also to reveal their actions. It is also possible to determine what a user is touching. Note that while information about who a user meets could also be considered behavioral, we assign it to \ref{itm:L1} because it primarily follows from location and movement information.
    \item[\namedlabel{itm:L3}{L3})] \emph{Physiological Data:} The most privacy-related challenge in sensing is the acquisition of physiological data, including breathing frequency and heart rate \cite{11416194,ZhaoAccess2024}. Despite the persisting complexity of these tasks, the advent of sophisticated hardware, such as multiple antennas, and software, including signal processing techniques and machine learning algorithms, has led to significant advancements. The imminent availability of higher radio frequencies, characterized by enhanced spatial resolution, further augments the potential of 6G networks to provide such information. Evidently, the disclosure of such information is indicative of health data and may be regarded as particularly critical from a privacy standpoint, as stipulated in the \ac{gdpr}.
\end{enumerate}

It might be possible to avoid some privacy issues by limiting the 6G network at deployment to a specific level of sensing capability. However, that may not always be possible or desirable, and other privacy-preserving solutions must be integrated within the network. We discuss some of these in Section~\ref{sec:solutions}. 

In the next sections, we use the aforementioned classification to structure our discussions on sensing applications, privacy challenges, and solutions. 

\section{Sensing Applications}\label{sec:applications}

Sensing is widely expected to provide efficiency, reliability, and contextual intelligence to the network by integrating and coordinating sensing capabilities and communication technologies~\cite{LiCuMaXuHaElBu22}. Sensing applications can be categorized according to their scope. \emph{Internal} applications use sensing to optimize or protect the operation of the network itself, whereas \emph{external} applications rely on sensing as an enabler for vertical services beyond network operation. In this work, we identify representative internal and external applications according to the privacy-sensitive sensing-data levels introduced in Section~\ref{sec:level_sensing_granularity}. 

\subsection{Internal Applications}
\label{sec:internal_apps}
Internal applications of ISAC envision improving network operation and performance, for example, through detection and adaptation of communication states and security threat classification/mitigation through network-based monitoring.

\subsubsection{Communications State Detection and Adaptation.} 
The utilisation of  ISAC can provide insights into the received signal, channel conditions, interference sources, and environmental factors. The real-time access to such information can be exploited to dynamically adapt the communication transmission parameters (e.g., waveform design, modulation scheme, bandwidth, and power allocation) on a link basis, or even to apply a cluster of \acp{RIS} in a coordinated manner to achieve end-to-end optimised connectivity~\cite{ LiKhAhSoKhShAsHa25}. This application mainly relies on \ref{itm:L1} data, because channel and environmental information may reveal location, movement, blockage, and occupancy-related information.

\subsubsection{Security Threat Classification and Mitigation.}

The accurate estimation of physical-layer features can support the detection and classification of attacks, with the aim of protecting communication and sensing functions against unauthorized entities~\cite{NaAlKaHuYu23,MaOcSh25,OnurSecureCommunicationComputationTutorial}. 

In the context of \ac{ISAC}, the network's sensing functionality can be used to monitor the radio environment for anomalous physical presences. For example, by analyzing high-resolution \ac{CSI} and Doppler shifts, the system may identify entities that do not transmit but alter the propagation environment, and adapt the beamforming strategy accordingly to adaptively null the beam in the direction of the intruder~\cite{IlQaAlAlAldRMaRoAmAK24}.

This application primarily relies on \ref{itm:L1} sensing, since environmental mapping and location-related information are used to detect anomalous objects or presences. Furthermore, recent advances in physical-layer security and \ac{ML}-based monitoring can support the real-time classification and mitigation of sophisticated wireless threats, such as pilot contamination and smart jamming, in ISAC-related settings~\cite{NaAlKaHuYu23,MaOcSh25,OnurSecureCommunicationComputationTutorial}. Thus, more advanced classification may also use features related to movement or device behavior, and therefore may involve \ref{itm:L2} data.

\subsection{External Applications}
External applications use sensed network information to enable vertical-specific services beyond network operation. Representative scenarios include environmental monitoring, living-environment monitoring and control, and management of mobile agents. For each application class, we indicate the relevant sensing-data levels, \ref{itm:L1}-\ref{itm:L3}, defined in Section~\ref{sec:level_sensing_granularity}.

\subsubsection{Environmental Monitoring (level~\ref{itm:L1}).}
\ac{ISAC} technology can be applied to environmental and climatic monitoring~\cite{3GPP_usecase,ETSI_sec,WuJiWaYaZhYiNa24}, including physical parameters such as temperature, humidity, and wind speed, as well as air or water quality, pollutant presence, and disaster monitoring, e.g., earthquakes, fires, and floods. In this work, this application is mainly associated with \ref{itm:L1}, because the sensed information concerns the physical environment rather than behavioral or physiological attributes.

\subsubsection{Living Environments Monitoring and Control (levels \ref{itm:L2} and \ref{itm:L3}).}
The improved precision of sensing capabilities in indoor scenarios can support intelligent living environments and human-to-machine interaction. In particular, by feeding the \ac{CSI} data or derived features to \ac{ML} models, \ac{CSI} sensing can enable the recognition of large- and small-scale movements, such as activities, e.g., walking, running, and jumping, or gestures~\cite{Meneghello2023,Li2022,Ma2018}, as well as person identification~\cite{martins2024wifi,Li2022}. Applications include sign-language translation~\cite{Ma2018}, human-machine interfaces in \ac{IoT} applications, activity monitoring, and physiological-signal monitoring~\cite{WuJiWaYaZhYiNa24}. These applications mainly involve \ref{itm:L2} data when activity, gesture, posture, or identity-related behavior is inferred, and \ref{itm:L3} data when physiological signals are monitored. 

\subsubsection{Mobile Applications (levels \ref{itm:L1} and \ref{itm:L2}).}\label{sec:mobile_apps} 

\ac{ISAC} is a key enabler for mobile and highly dynamic scenarios in future \ac{6G} systems, where rapid environmental changes require coordination between perception and data exchange. In such settings, \ac{ISAC} allows network nodes to infer spatial context while maintaining connectivity, supporting mobility-aware decision making for autonomous and connected agents~\cite{LiCuMaXuHaElBu22,OnurChalmersISACtutorial}. Specifically, for vehicular networks, ISAC enables the joint detection of the operating environment and neighboring mobile devices. This synergy supports advanced \ac{V2X} communication schemes, for example, in autonomous-agent platooning, where real-time sensing of surrounding gaps and velocities is important for stability and safety under dynamic traffic conditions~\cite{LiGuHeMYuZhGu24}. Beyond road traffic, similar principles extend to coordinated operation of mobile agents such as unmanned aerial vehicles, where groups of mobile nodes can adapt their trajectories, avoid obstacles, and maintain formation without relying exclusively on external sensors~\cite{GPFuKaVaDaShShByWy24}. The data gathered in these scenarios typically maps to \ref{itm:L1} for environmental mapping and trajectory tracking, and may involve \ref{itm:L2} when mobile objects, users, or hazards are characterized beyond their location and movement.

\subsubsection{Industrial Site Coordination (levels \ref{itm:L1} and \ref{itm:L2}).}
The distributed integration of communication and dynamic context-sensing capabilities among nodes can support collaboration in \ac{6G} intelligent factories~\cite{GPFuKaVaDaShShByWy24}. This can improve the coordination of mobile agents, including workers, vehicles, robots, and cobots, and consequently enhance overall operational efficiency. In particular, mobility can be managed by using \ac{ISAC} for trajectory tracking and hazard or obstacle detection in workplace-safety applications. Moreover, the inherent high data rates can support industrial \ac{VR}/\ac{AR} applications, where behavioral monitoring may enable immersive training, design, and maintenance tasks. Finally, \ac{ISAC} can support closed-loop and real-time monitoring and control of industrial processes, including spatial and differentiated access control and predictive equipment maintenance. In these scenarios, \ref{itm:L1} data arises from tracking, mapping, occupancy, and obstacle detection, whereas \ref{itm:L2} data may arise when worker behavior, gestures, posture, or interaction patterns are inferred.

Emerging sensing-based applications, either internal or external, span sensing granularities from location and environment data \ref{itm:L1} to behavioral data \ref{itm:L2} and, in some cases, physiological data \ref{itm:L3}. These levels are useful for organizing privacy analysis because the same sensing information that enables network optimization, mobility support, living-environment control, or industrial coordination may also reveal information about users, devices, bystanders, passive objects, and environments. The following section, therefore, discusses the corresponding privacy challenges.

\section{Privacy Challenges} \label{sec:challenges}

This section identifies the privacy challenges that follow from sensing-derived personal and environmental information in \ac{ISAC}. We first recall the relevant scope of personal data and profiling under the \ac{gdpr}. We then discuss governance challenges in sensing-data handling, followed by risks associated with the three sensing-data levels introduced in Section~\ref{sec:level_sensing_granularity}. Note that in this section, the adversary can be of different types: from an honest-but-curious network operator to a rogue base station, from an unauthorized third-party application to a malicious bystander \ac{UE} sniffing reflected waves.

\subsection{Definitions and Scope}

As specified in Section~\ref{sec:introduction}, this paper considers the disclosure or loss of control over sensing-derived personal or environmental information as a privacy breach, whereas the disclosure of message content is treated as a confidentiality and security issue. Not only active users, but also bystanders and passive objects in the target sensing area may be sensed, which can lead to the collection of sensitive contextual information~\cite{ETSI_sec}. To provide a more precise definition of the concept of privacy, we turn to the most recent literature on the subject, particularly the \ac{gdpr} of the European Union. This regulatory instrument has established de facto global standards regarding privacy, defining {\em personal data} as \cite[Art. 4]{gdpr} "any information relating to an identified or identifiable natural person (‘data subject’); an identifiable natural person is one who can be identified, directly or indirectly, in particular by reference to an identifier such as a name, an identification number, location data, an online identifier or to one or more factors specific to the physical, physiological, genetic, mental, economic, cultural or social identity of that natural person".

\ac{ISAC} can facilitate the acquisition of location data, movement-related information, and, in some applications, health-related information. Such data may support profiling, a concept further delineated in the \ac{gdpr}. According again to  \cite[Recital 71]{gdpr}, profiling "consists of any form of automated processing of personal data evaluating the personal aspects relating to a natural person, in particular to analyze or predict aspects concerning the data subject’s performance at work, economic situation, health, personal preferences or interests, reliability or behavior, location or movements, where it produces legal effects concerning him or her or similarly significantly affects him or her." According to the \ac{gdpr}, the practice of profiling is permissible only under specific conditions. Where profiling or solely automated decision-making has legal or similarly significant effects, \ac{gdpr} imposes specific conditions and safeguards, which may include explicit consent or other legally recognized bases. In all cases, the use of personal data must be clearly delimited and supported by appropriate privacy-aware safeguards. For further insights into the legal aspects of  \ac{ISAC} see also \cite{report-CNIT}. Consequently, \ac{ISAC} design should account for privacy requirements from the outset, including limits on sensing purpose, data access, processing, storage, and disclosure. In this paper, we focus on the technical and architectural implications of these requirements rather than providing a legal analysis. 

The European standardisation work on \ac{ISAC} has examined the privacy issues and proposed new potential requirements to address these challenges in \ac{ISAC}-enabled \ac{6G} systems \cite{ETSI_sec}. On the other hand, several papers have also surveyed the privacy issues of \ac{ISAC} in \ac{6G} networks \cite{qu2026secure,10608156, su2025integratingsensingcommunications6g,11015430, dassUEpriv}, focusing on technical solutions to protect some aspects (such as location privacy or the leakage of the sensing signal to unauthorized receivers), rarely addressing the main and fundamental shortfalls of \ac{ISAC} with respect to privacy. The Commonwealth Scientific and Industrial Research Organisation (CSIRO) has also issued a report \cite{suzuki2025security} on security and privacy issues in \ac{6G}, including a discussion on \ac{ISAC}. In contrast, this section organizes the privacy challenges according to governance requirements and the three sensing-data levels used throughout this work.

\subsection{Governance Challenges in Sensing-Data Handling}

\subsubsection{Consent and Authorization for Sensing.}

In conventional communication systems, tracking is mainly associated with active and connected \acp{UE}. In contrast, \ac{ISAC} extends the scope of observation to the surrounding physical environment, where sensing may lead to collateral data collection involving bystanders and passive objects. As a result, \ac{ISAC} may enable the collection of \ac{PII} related not only to connected users, but also to individuals who do not carry or use a \ac{UE} within the sensing environment~\cite{3GPP_usecase, ETSI_sec}.

Use cases involving 3GPP sensing in public areas, such as intruder detection on highways and railway crossings, tourist-spot traffic management, and UAV-based applications, may therefore involve the collection of PII related to human beings. In restricted or sensitive areas, such as military zones, sensing may further expose sensitive information, including geolocation data, operational activities, and infrastructure details. Therefore, where applicable and possible, consent or authorization from public management bodies and relevant authorities should be obtained before enabling 3GPP-based sensing services. In addition, when the network intends to use \acp{UE} as sensing entities or sensing-data providers, explicit user consent and appropriate privacy controls are required before such involvement~\cite{3GPP_usecase, dass2024addressing, dassUEpriv}.

\subsubsection{Transparency, Purpose Limitation, and User Control.}

Depending on the application request, the network may collect sensing data from an area of interest and may also involve \acp{UE} in the sensing process. However, the purpose of sensing, the intended use of the sensing data, the processing entities involved, and the parties authorized to access the sensing results may not always be transparent to users or \ac{UE} owners. In such cases, sensing may occur without sufficient authorization, purpose limitation, or user control.

For example, sensing data collected through \ac{UE}s for pedestrian detection at crossroads may later be repurposed for commercial analytics, such as estimating traffic density near shopping malls. Similarly, users may be unaware that movement-related information collected to improve traffic flow is also being used to support commercial decision-making or marketing strategies. Beyond the original sensing purpose, the disclosure of sensing results to third parties may also reveal PII related to users, \ac{UE}s, and bystanders present in the sensing environment.

\subsection{Location and Environmental Privacy Risks}

The first sensing-data level \ref{itm:L1} concerns location and environment data. A central privacy risk at this level is the ability of \ac{ISAC} to localize and track moving or static objects. Different from localization techniques that are able to obtain the position of active transmitters (e.g., through triangulation and or trilateration), the sensing capabilities of \ac{ISAC} enable the positioning of passive objects, and these include also people. In this case, communication signals may also be used for radar-like sensing. Such sensing can be difficult for affected individuals to detect, making transparency and control challenging.  

\ac{ISAC} systems have the capability to capture location data of users at varying levels of granularity, ranging from broader zones or areas to precise coordinates. There are different methods, such as \ac{TDoA}, \ac{AoA}, \ac{AoD} measurements, \ac{RSSI}, etc., to detect and locate static and moving objects \cite{3GPP_usecase, JCAS_survey}. The network operator can also combine location estimates over time to infer location histories. Base stations with sensing capabilities may also cooperatively track movement. In \ac{ISAC} applications, such as pedestrian detection at crossroads and automotive maneuvering, the location of the users requesting the service is continuously monitored to identify any obstacles around them \cite{3GPP_usecase}.

\ac{ISAC} can also be used to determine occupancy of spaces and, when sensing features become more detailed, to infer activity or identity-related characteristics~\cite{Iker_Counting,GaitRecon}. These capabilities create a boundary between \ref{itm:L1} risks, such as occupancy and presence detection, and \ref{itm:L2} risks, such as activity or identity-related behavioral inference, including sensing through opaque obstacles~\cite{adib2013see}.

\subsection{Behavioral Profiling Risks}
The second sensing-data level \ref{itm:L2} concerns behavioral data. From sensing data captured in the environment, sensing units or network operators may infer behavior of users, \acp{UE}, or other objects in the sensing area~\cite{radar_profiling}.

While behavioral inference in \ac{ISAC} applications can support user experience and health-related monitoring, it also poses significant privacy risks. For instance, it can enable surveillance of routine behaviors, including where users move, when they interact with others, and contextual information such as walking or sleeping habits, by observing phase and/or amplitude patterns of the \ac{CSI}~\cite{10355064, WLAN_sensing}. Moreover, an adversary with access to \ac{ISAC} sensing signals or reflected signals may analyze \ac{CSI} patterns to infer behavioral properties of objects in the sensing area. Behavioral profiling may also concern the sensing process itself, including when sensing signals are transmitted, which area is sensed, and which \acp{UE} are involved. If sensing requests are exposed, an adversary may further link applications to the types of sensing data they request.

\subsection{Physiological and Biometric Privacy Risks}

The third sensing-data level \ref{itm:L3} concerns physiological and biometric information. \ac{ISAC} systems may support the extraction of sensitive personal attributes from the sensed environment, including physical dimensions, biometric features, and physiological indicators. In applications such as domestic health and sleep monitoring, radar-based sensing can facilitate the estimation of vital signs, such as heart rate and respiration, and may also support biometric identification~\cite{radar_biometric, radar_signature}. Consequently, unauthorized sensing or misuse of \ac{ISAC} infrastructure may expose personal attributes of humans or other targets within the sensing range. Such exposure of \ac{PII} creates significant privacy risks, including identifiability and linkability threats~\cite{LINDDUN, dass2024addressing}.

Note that under current frameworks (like the European Union \ac{AI} Act), if a \ac{6G} base station processes radio data to classify human behavior (L2) or infer physiological vital signs (L3), the telecommunications operator is no longer a mere data conduit and it may become the operator of a high-risk biometric categorization \ac{AI} system. This implies severe liabilities if ``always-on" sensing occurs without a legal mandate.

\section{Solutions to Privacy Challenges}
\label{sec:solutions}

Section~\ref{sec:challenges} identified privacy risks that arise when sensing-derived information is collected, inferred, or disclosed. This section reviews representative solution directions and maps them to the three sensing-data levels introduced in Section~\ref{sec:level_sensing_granularity}. The emphasis is not on communication-message confidentiality, but on mechanisms that limit spatial observability at L1, restrict behavioral inference at L2, and control the collection, processing, and disclosure of highly-sensitive physiological or biometric information at L3.

\subsection{L1: Location \& Environment Privacy}
\label{subsec:loc}

At L1, privacy protection mainly concerns spatial observability: whether positions, movements, occupancy, or environmental structure can be inferred, and at what resolution. The representative mechanisms below therefore aim to reduce, obfuscate, or selectively control the spatial information available to unauthorized sensing receivers through signal processing and physical-layer techniques.

\paragraph*{Location Obfuscation:}
Radio-based positioning techniques can achieve sub-meter localization accuracy~\cite{signals4010006}, which creates privacy risks when users, bystanders, or passive objects can be localized without meaningful control. Location obfuscation addresses this risk by deliberately reducing the precision or reliability of the location information exposed by the sensing system. 

For example, Nguyen et~al.~\cite{10355064} propose mitigation strategies to control the precision, exposure, and usability of location information in sensing systems. They advocate privacy-aware positioning, where spatial and temporal resolution is deliberately limited based on application needs. Proposed solutions include location obfuscation, uncertainty-aware reporting, and access control mechanisms that restrict who can query location data and at what precision, supported by context-aware policies and user-centric privacy controls.

Keya et~al.~\cite{10773968} propose a solution for \ac{ISAC} systems based on channel obfuscation, where privacy is achieved by deliberately manipulating the wireless channel so that true user locations become indistinguishable from a predefined set of alternative locations. Specifically, the core idea is to control the resolution of location detection by transforming the \ac{CFR} such that the sensed signal associated with a true location becomes statistically close to signals corresponding to multiple ``dummy'' locations (an obfuscation set), increasing uncertainty for both passive and active attackers.

\paragraph*{Artificial Noise Injection:}
Artificial-noise-based designs protect L1 privacy by degrading unauthorized sensing while preserving the intended sensing and communication functions. In this case, privacy is addressed by controlling the sensing information available to an unauthorized receiver. 

Zou et~al~\cite{10587082} consider the problem that the same dual-functional waveform used for legitimate sensing and communication can also be exploited by unauthorized receivers (“sensing Eves”) to infer environmental and location information. The proposed solution is the introduction of a sensing-security framework based on beamforming and \ac{AN} injection, where sensing performance is intentionally controlled rather than universally maximized. The sensing mutual information of the legitimate sensing receiver is maximized while constraining the mutual information obtainable by an eavesdropper. To further suppress unauthorized sensing, the system injects artificial noise into the transmitted waveform, degrading the sensing quality and estimation capability of the Eve without significantly affecting the intended sensing and communication tasks.

\paragraph*{\ac{RIS} Based Approaches:}
RIS-assisted \ac{ISAC} privacy approaches can actively manipulate the propagation environment to obfuscate sensing targets and reduce unauthorized localization accuracy. 

The work by Magbool et~al.~\cite{magbool2025hidingplainsightrisaided} studies how \ac{RIS} can be used to actively protect sensing privacy by reshaping the wireless environment. Instead of allowing all receivers to exploit high-resolution sensing signals, the \ac{RIS} is configured to degrade the sensing quality at unauthorized observers, making targets harder to detect or localize, while still maintaining acceptable performance for legitimate users. This enables a form of selective sensing visibility, where spatial information, such as location and presence, becomes unreliable for adversaries. From a privacy perspective, the key benefit is that sensing accuracy itself becomes a controllable resource, allowing ISAC systems to deliberately introduce uncertainty into environmental and localization data without disabling sensing services altogether.

\paragraph*{Directional Sensing Constraints:}
Directional constraints protect L1 privacy by making angular information less useful for localization while maintaining the communication link performance. 

Khan et~al.~\cite{khan2025beamformingtransmitterlocationprivacy} propose a beamforming strategy that protects user location privacy by manipulating the angular characteristics of transmitted signals so that the true direction of arrival cannot be reliably inferred by a sensing receiver. Instead of simply reducing signal power, the method introduces a \ac{DoA-OPR} constraint, which ensures that misleading or false angular peaks dominate over the true direction in the received signal. By jointly optimizing communication performance and this obfuscation constraint, the approach allows the system to maintain reliable communication while distorting directional sensing information, effectively preventing accurate localization.

Ma et~al~\cite{10615848} propose a physical-layer privacy mechanism that limits the ability of \ac{ISAC} receivers to infer a transmitter’s location by constraining the directional characteristics of the transmitted signal. Instead of maximizing beamforming gain toward a specific direction, the authors design a sensing-resistance beamformer that spreads or reshapes the angular power distribution to make the true direction less distinguishable.

\subsection{L2: Behavioral Privacy (Activity, Gait, Gesture)}
\label{subsec:behav}

At L2, privacy protection concerns the inference of activity, gait, gesture, posture, identity-related behavior, or other behavioral patterns from sensing-derived data. The relevant solution directions therefore move beyond spatial obfuscation and focus on controlling the features, representations, model updates, or functions that are exposed to applications or other network entities. Emerging directions to address this are data-driven approaches that learn privacy-preserving representations, distributed and federated learning, and information-theoretic approaches that explicitly constrain the functions computed from sensing data.

\paragraph*{ML-Based Privacy-Preserving Representation Learning:}
Privacy-preserving \ac{ML} techniques, including adversarial learning, representation learning, and federated learning, are relevant for limiting sensitive inference in \ac{6G} systems~\cite{9146540}. In the present L2 context, such methods can be used to preserve task-relevant information, such as activity recognition, while reducing exposure of identity-related or behavioral attributes.

An example of this principle is given by Delgado-Santos et~al.~\cite{DELGADOSANTOS202230}, proposing an approach based on autoencoders and adversarial training to transform gait data into privacy-preserving representations. The method preserves the utility of gait signals for authentication while suppressing sensitive attributes such as gender or contextual behavioral information, demonstrating how feature-level sanitization can mitigate behavioral privacy risks in sensing systems.

\paragraph{Distributed and Federated Learning:}
A complementary approach to protecting behavioral privacy in \ac{ISAC} systems is to avoid centralized collection of behavioral sensing data. In distributed or federated learning architectures, raw sensing data can remain at the device or network edge, while only model updates or aggregated parameters are shared. This can reduce the exposure of activity, gait, or gesture information to centralized entities.

For example, Zhang et~al.~\cite{9626548} propose a federated-learning-based framework for WiFi \ac{CSI} activity recognition that enables collaborative training across multiple environments without sharing raw sensing data. The proposed approach keeps local sensing data on-device or within the local environment, while only sharing learned model parameters for global training. Behavioral inference models can be collaboratively trained while limiting direct exposure of sensing data, reducing the risk of centralized behavioral profiling and unintended information leakage.

\paragraph*{Privacy-preserving feature extraction and randomized inference:}
A further L2 direction is to restrict the behavioral functions computed from \ac{CSI}, radar, or localization traces, instead of trying to protect raw measurements or unnecessarily detailed intermediate features. This is aligned with the privacy notion used in this paper, where the concern is loss of control over sensing-derived information that enables activity, gait, gesture, or posture inference.

Randomized function-computation models provide related conceptual tools for this direction. The \ac{RDFC} framework studies the synthesis of randomized functions of distributed data under privacy constraints~\cite{OnurRDFC,OnurDeepRDFC}, while private remote-source function computation studies tradeoffs between function computation and information leakage for remote sources, such as semantics of data ~\cite{OnurPrivateFunctionComp}. These works provide useful abstractions for future sensing pipelines in which applications receive only controlled, possibly randomized, sensing-derived outputs rather than full signal representations.

\subsection{L3: Physiological Privacy (Vital Signs, Biometrics)}\label{subsec:phys}
At L3, the privacy concern is the exposure of physiological or biometric information, such as breathing patterns, heart-rate-related features, or biometric identifiers. Compared with L1 and L2, protection at this level requires stricter data minimization and lifecycle control, because the information may be sensitive, persistent, and difficult to revoke once disclosed. The main design objective is therefore to limit collection where possible and, when collection is necessary, to tightly control processing, storage, access, and disclosure within \ac{ISAC} systems.

The work by Dass et~al.~\cite{DassZK25} proposes a privacy-preserving \ac{ISAC} architecture in which sensing, processing, and data access are separated into controlled functional components. The architecture supports policy-based access control, local processing, and purpose-limited disclosure for sensing requests handled by the network. Although the work does not focus on physiological sensing, it is particularly relevant at L3 because physiological or biometric sensing requires strict control over what is collected, where it is processed, and which applications may access the resulting information.

\section{Research Roadmap}

The preceding sections show that privacy in \ac{ISAC} concerns the collection, processing, inference, storage, and disclosure of sensing-derived information about active users, \acp{UE}, bystanders, passive objects, and environments. Based on the sensing-data levels in Section~\ref{sec:level_sensing_granularity}, the applications in Section~\ref{sec:applications}, the privacy challenges in Section~\ref{sec:challenges}, and the solution directions in Section~\ref{sec:solutions}, the following research directions are central for privacy-preserving \ac{ISAC}.

\begin{enumerate}
\item \textbf{Privacy-by-design for sensing signals and sensing capabilities.}
Future \ac{ISAC} systems require designs in which privacy leakage is considered together with communication rate, sensing accuracy, latency, and energy consumption. Key open problems are to define privacy metrics for the sensing-data levels \ref{itm:L1}-\ref{itm:L3}, adapt waveforms and beams to the sensing resolution required by each service, and limit unauthorized sensing.
\item \textbf{Location and environment privacy under high-resolution sensing.}
Location and environment data form the first privacy-sensitive layer of \ac{ISAC}. Future work must protect active users, bystanders, passive objects, and environments in the sensing area. Important directions include pseudolocation, location obfuscation, privacy-aware beam and access-point selection, and service-level rules that avoid unnecessary location precision.
\item \textbf{Data minimization and privacy-preserving sensing pipelines.}
Privacy protection should be integrated before raw \ac{CSI}, radar, and localization data are stored, transferred, or disclosed. Open problems include identifying application-specific minimal sensing features, separating raw measurements from application-facing results, enforcing retention limits, and combining anonymization, pseudonymization, encryption, and access control in \ac{ISAC} functions.
\item \textbf{Protection against behavioral and physiological inference.}
Behavioral and physiological sensing correspond to the more privacy-sensitive levels \ref{itm:L2} and \ref{itm:L3}. Future \ac{ISAC} research must limit unnecessary behavioral and physiological inference while preserving the intended sensing service. Key directions include feature sanitization for \ac{CSI} and radar data, privacy-preserving activity recognition, and privacy-aware handling of biometric or health-related sensing results.
\item \textbf{Consent, transparency, data ownership, and accountability.}
Technical privacy mechanisms must be supported by governance mechanisms for sensing requests and sensing results. Future architectures need explicit sensing policies, consent management, transparency logs, and fine-grained authorization. Open problems include data ownership for bystander sensing, permission revocation, third-party auditing, and measurable privacy criteria for system design.
\end{enumerate}

These directions are interdependent. Physical-layer mechanisms influence what can be sensed, data-minimization mechanisms influence what is stored or shared, learning-based pipelines influence what can be inferred, and governance mechanisms determine who may request, process, and disclose sensing results. Addressing these aspects together is necessary for \ac{ISAC} systems that provide useful sensing services while respecting privacy expectations and legal requirements.

\section{Conclusion}

In this work, we discussed privacy challenges and solution directions for next-generation \ac{ISAC} systems. We focused on privacy issues caused by sensing-derived information, rather than the confidentiality of communication content. In particular, we emphasized that \ac{ISAC} can affect not only active users and their \acp{UE}, but also bystanders, passive objects, and the surrounding environment. To organize these risks, we classified privacy-sensitive sensing data into three levels: location and environment data, behavioral data, and physiological data. This classification was then used to relate sensing capabilities and applications to privacy concerns.

We summarized representative internal and external \ac{ISAC} applications and showed that their privacy impact depends on the type and granularity of the sensing data they require. We then reviewed privacy challenges related to consent, transparency, data ownership, profiling, location tracking, behavioral inference, and physiological sensing. Representative solution directions were discussed, including control of spatial observability, location obfuscation, \ac{AN}-based protection, \ac{RIS}-based privacy approaches, directional sensing constraints, privacy-aware system architectures, and privacy-preserving feature extraction mechanisms.

Finally, we outlined a research roadmap for privacy-preserving \ac{ISAC}. The main open directions include privacy-by-design for sensing signals and sensing capabilities, location and environment privacy under high-resolution sensing, data minimization in sensing pipelines, protection against unnecessary behavioral and physiological inference, and governance mechanisms for consent, transparency, ownership, and accountability. Addressing these directions together is necessary for future \ac{ISAC} systems that provide useful sensing services while respecting privacy expectations and legal requirements.

\begin{credits}

\subsubsection{\ackname} The work of O. Günlü was partially supported by the ZENITH Research and Leadership Career Development Fund under Grant ID23.01, EU COST Action 6G-PHYSEC, Swedish Foundation for Strategic Research (SSF) under Grant ID24-0087, and German Federal Ministry of Research, Technology and Space (BMFTR) 6GEM+ Transfer Hub under Grants 16KIS2412 and 16KISS005. The work of P. Dass is financed based on the budget passed by the Saxonian State Parliament in Germany. The work of J.P.~Vilela is funded by national funds through FCT – Fundação para a Ciência e a Tecnologia, I.P., under the support UID/50014/2025 (\url{https://doi.org/10.54499/UID/50014/2025}).

\subsubsection{\discintname}
The authors have no competing interests to declare that are relevant to the content of this article. 
\end{credits}

\bibliographystyle{splncs04}
\bibliography{mybibliography}

\end{document}